\documentclass[aps,prc,showpacs,twocolumn,superscriptaddress]{revtex4}
\usepackage{multirow}
\usepackage{graphicx}
\usepackage{amsmath}
\usepackage{mathrsfs}
\usepackage{ulem}

\def\version{version}

\newcommand{ \be }{\begin{equation}}
\newcommand{ \ee }{\end{equation}}
\newcommand{ \bea }{\begin{eqnarray}}
\newcommand{ \eea }{\end{eqnarray}}

\usepackage{color}

\begin{document}
\title{
\begin{flushright}
{
\small \sl \version 1.3\\
}
\end{flushright}

Eccentricity is not the only source of elliptic flow fluctuations

}

\author{Kai Xiao}\affiliation{Key Laboratory of Quark and Lepton Physics (MOE) and Institute of Particle Physics, Central China Normal University, Wuhan 430079, China}

\author{Feng Liu}
\affiliation{Key Laboratory of Quark and Lepton Physics (MOE) and Institute of Particle Physics, Central China Normal University, Wuhan 430079, China}

\author{Fuqiang Wang}\email{fqwang@purdue.edu}
\affiliation{Key Laboratory of Quark and Lepton Physics (MOE) and Institute of Particle Physics, Central China Normal University, Wuhan 430079, China}
\affiliation{Department of Physics, Purdue University, West Lafayette, Indiana 47907, USA}

\date{\today}

\begin{abstract}
Sources of event-by-event elliptic flow fluctuations in relativistic heavy-ion collisions are investigated in a multiphase transport model. Besides the well-known initial eccentricity fluctuations, several other sources of dynamical fluctuations are identified. One is fluctuations in initial parton configurations at a given eccentricity. Second is quantum fluctuations in parton interactions during system evolution. Third is fluctuations caused by hadronization and final-state hadronic scatterings. The magnitudes of these fluctuations are investigated relative to eccentricity fluctuations and average flow magnitude. The fluctuations from the latter two sources are found to be negative. The results may have important implications to the interpretation of elliptic flow data.

\end{abstract}

\maketitle
\clearpage

{\bf Introduction}
A strongly interacting quark-gluon plasma (sQGP) is created in relativistic heavy-ion collisions~\cite{whitepapers}.  At nonzero impact parameter, the transverse overlap region of colliding nuclei is anisotropic. Due to interactions among constituents, the produced matter undergoes a rapid anisotropic expansion resulting in an anisotropic distribution of final-state hadrons in momentum~\cite{Ollitrault}. This anisotropy can be quantified by the second coefficient ($v_{2}$) of the Fourier expansion of the final-state particle azimuthal distribution, called elliptic flow~\cite{Voloshin1}.  Because the anisotropy in configuration space is quickly diminished due to anisotropic expansion, elliptic flow is primarily sensitive to the early stage of sQGP evolution. Shear viscosity is known to damp the development of anisotropic flow. Anisotropic flow data in comparison to hydrodynamical calculations may, therefore, measure the shear viscosity to entropy density ratio ($\eta/s$)~\cite{Heinz}.

Elliptic flow has been extensively studied in relativistic heavy-ion collisions~\cite{whitepapers}. Because the initial-state geometry is not experimentally accessible, anisotropic flow is often measured by two-particle correlations~\cite{Poskanzer}. The measured quantity is the root mean square, $\sqrt{\langle v_{2}^{2} \rangle}$. Event-by-event flow fluctuations are therefore critical and the understanding of these fluctuations is essential in extracting physics information from flow measurements.

At a given impact parameter ($b$), the interacting nucleons are not identically distributed due to fluctuations in nucleon distribution in a nucleus and due to the quantum nature of nucleon-nucleon interactions. As a consequence, elliptic flow develops relative to the so-called participant plane~\cite{PHOBOS1}, not the reaction plane defined by the beam and impact parameter directions. It is believed that flow fluctuations are determined by fluctuations in the initial-state geometry anisotropy~\cite{PHOBOS2}. The initial-state anisotropy is often quantified by eccentricity. In many flow studies, the final-state anisotropy is assumed to be strictly proportional to the eccentricity. This strict proportionality, under the assumption of Gaussian fluctuations in the $x$ and $y$ components of eccentricity, leads to a well-defined Bessel-Gaussian in the final-state flow parameter~\cite{Voloshin2}. However, at a given eccentricity and impact parameter, there can still be fluctuating distributions of interacting nucleons. Are these fluctuations important to elliptic flow development? This paper tries to answer this question.

Given a fixed initial condition (eccentricity and configuration), the evolution of hydrodynamics is determined. The final state anisotropy from hydrodynamical calculations is fixed. However, there may be other sources of dynamical fluctuations during the stage of system evolution, e.g. parton-parton interactions and hadronic scatterings, which could lead to additional flow fluctuations. This paper further attempts to address this question by employing the AMPT (A Multi-Phase Transport) model, because transport models inherit all quantum fluctuations in the interactions among constituents.

\begin{figure*}[htb]
\renewcommand{\figurename}{FIG.}
\vskip -0.4cm
\centerline{\includegraphics[width=0.85\textwidth]{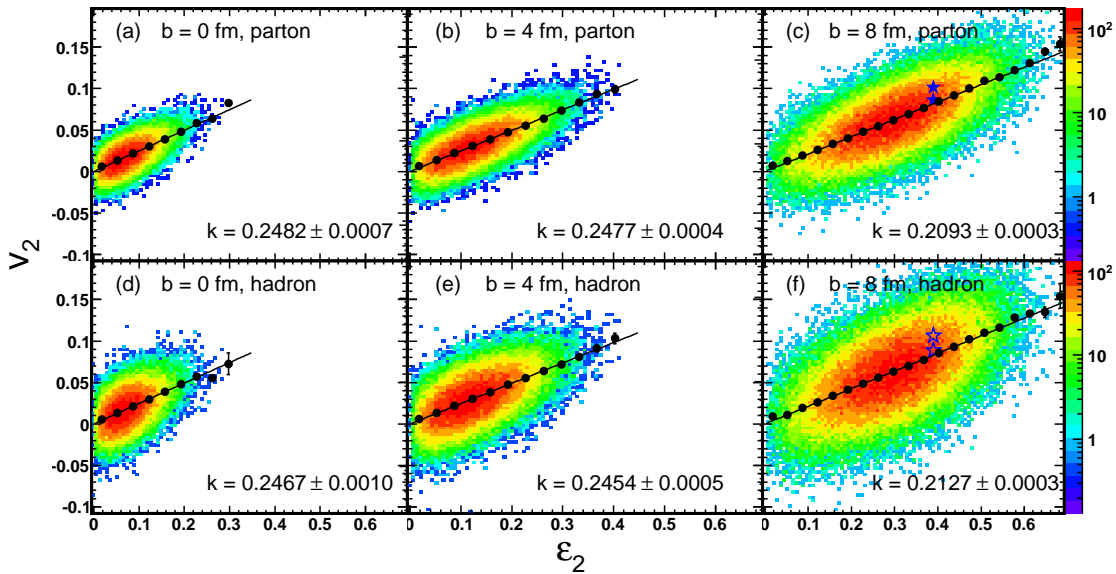}}
\vspace{-1.2em}
\caption{(Color online) Event-by-event correlations between momentum anisotropy $v_{2}$ and initial configuration eccentricity $\varepsilon_{2}$ for fixed impact parameter ($b=0$~fm, 4~fm and 8~fm) Au+Au collisions at $\sqrt{s_{NN}}$ = 200 GeV by the AMPT model (3 mb parton cross section). Upper and lower panels show $v_{2}$ of partons and charged hadrons, respectively, both within a pseudo-rapidity of $|\eta| < 1.0$. The solid dots show the average behavior of $\langle v_{2} \rangle$ versus $\varepsilon_{2}$ and the lines are one-parameter fits to $v_{2}$= $k \times \varepsilon_{2}$.  See text for the explanation of the four pentagram points in panels (c,f).}
\vspace{-1em}
\end{figure*}

{\bf Analysis Method}
AMPT~\cite{AMPT} consists of four main parts: the initial condition, parton-parton interactions, hadronization, and hadronic scatterings. The initial condition is obtained from the HIJING model~\cite{Wang}, which includes the spatial and momentum information of minijet partons from hard processes and strings from soft processes. The time evolution of partons is then treated according to the ZPC parton cascade model~\cite{Zhang}. After parton interactions cease, a combined coalescence and string fragmentation model is used for the hadronization of partons. The subsequent scatterings among the resulting hadrons are described by the ART model~\cite{Li} which includes both elastic and inelastic scatterings.

The initial geometric anisotropy of the transverse overlap region of a heavy-ion collision is often described by eccentricity~\cite{Alver}:
\begin{equation}
\varepsilon_{n} = \sqrt{\langle r^{2}\cos(n\phi_{part}) \rangle^{2} + \langle r^{2}\sin(n\phi_{part})\rangle^{2}}/\langle r^{2} \rangle
\end{equation}
where $r$ and $\phi_{part}$ are the polar coordinate positions of each parton liberated by the initial encounter of the colliding nuclei. As the system evolves, the initial configuration anisotropy is transferred to the momentum anisotropy by the hydrodynamical pressure gradient.

The momentum anisotropy is widely characterized by the Fourier coefficients~\cite{Poskanzer}:
\begin{equation}
v_{n} = \langle\cos[n(\phi - \Psi_{n})] \rangle,
\end{equation}
where $\phi$ is the particle azimuthal angle and $\Psi_{n}$ is the $n^{\rm th}$ harmonic plane angle. In AMPT $\Psi_{n}$ can be calculated in coordinate space by~\cite{Alver}

\begin{equation}
\Psi_{n}^{r} = ({\rm atan2}(\langle r^{2}\sin(n\phi_{part}) \rangle, \langle r^{2}\cos(n\phi_{part}) \rangle) + \pi)/n,
\end{equation} Note $\Psi_{n}^{r}$ is not necessarily the reaction plane due to event-by-event fluctuations. Due to the finite multiplicity of constituents, the constructed harmonic plane is smeared from the true one--the geometry harmonic plane of the participant partons in configuration space in the limit of infinite parton multiplicity--by a resolution factor. The resolution factor is calculated with an iterative procedure by the subevent method, dividing the constituents randomly into two subevents~\cite{Poskanzer}. Because of the large initial parton multiplicity, the calculated resolution is nearly unity~\cite{Xiao}.


Experimentally, the configuration space harmonic plane is inaccessible. The event plane reconstructed from final-state particle momenta is used as a proxy. Because of this, and neglecting nonflow, the measured anisotropic flow is its root mean square; it contains all fluctuation effects.




{\bf Results and Discussions}

\begin{figure*}[htb]
\hskip 0cm
\vskip -0.45cm
\centerline{\includegraphics[width=0.85\textwidth]{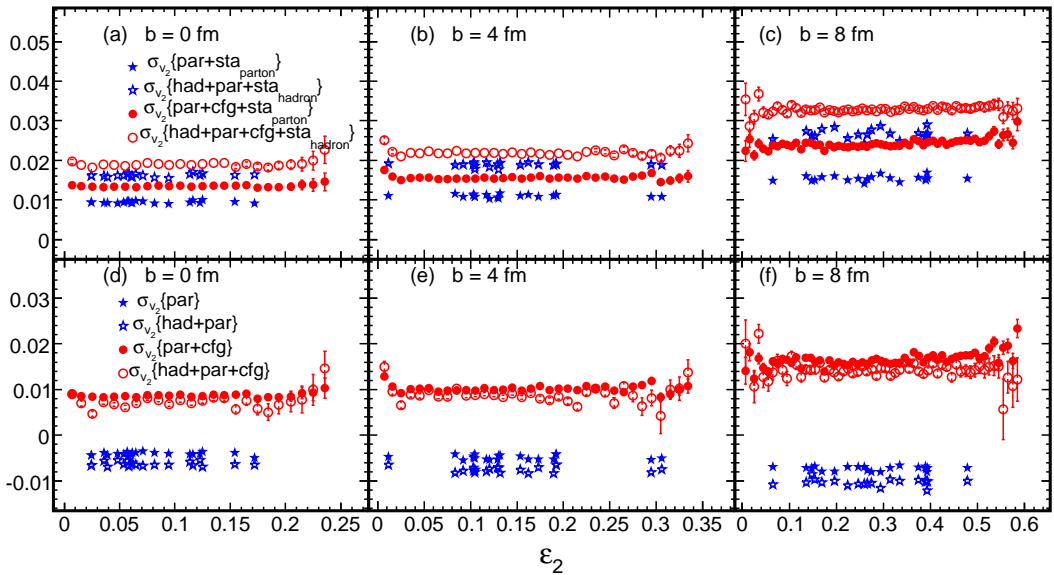}}
\vspace{-4em}
\caption{(Color online) $v_{2}$ fluctuations versus $\varepsilon_{2}$, before (upper panels) and after (lower panels) removal of statistical fluctuations. Red points are from default settings (from Fig. 1), while the blue points are from runs with identical parton configuration (examples shown in pentagrams in Fig. 1(c,f)).} \vspace{-1em}
\end{figure*}

Figure 1(a-c) show the event-by-event correlations between $v_{2}$ of partons after parton interactions cease and $\varepsilon_{2}$ from collisions at three fixed impact parameters. Figure 1(d-f) show those between $v_{2}$ of final-state charged hadrons after hadronic scatterings and $\varepsilon_{2}$. Before studying $v_{2}$ fluctuations, it is useful to first examine the average behavior of $\langle v_{2} \rangle$ vs. $\varepsilon_{2}$. This is shown by the solid dots. As the $\varepsilon_{2}$ increases, the magnitudes of average $\langle v_{2} \rangle$ increase linearly. The conversion power ($k$), the slope of $\langle v_{2} \rangle$ vs. $\varepsilon_{2}$ from a linear fit to the data, appears compatible between $b$ = 0 and 4 fm and smaller for $b$ = 8 fm. Interestingly, at a given $b$, almost the same k is observed for parton and hadron $v_{2}$ vs. $\varepsilon_{2}$. This indicates that the final-stage hadronic scatterings in AMPT do not generate significant additional $\langle v_{2} \rangle$.

Large fluctuations in $\varepsilon_{2}$ are observed in Fig. 1, which are due to geometry fluctuations at a fixed $b$. With increased $b$, $\varepsilon_{2}$ fluctuations are larger.  For a given $\varepsilon_{2}$, however, there still exist wide dispersions in $v_{2}$. This indicates that $v_{2}$ fluctuations are not solely due to $\varepsilon_{2}$ fluctuations; There are additional fluctuation sources in $v_{2}$. One source is simply statistical fluctuations (and they are larger in hadron $v_{2}$ than parton $v_{2}$ due to the smaller number of hadrons than partons). There may be sources of dynamical fluctuations. We first investigate fluctuations in initial parton configurations at fixed eccentricity. The same $\varepsilon_{2}$ does not necessarily mean the same initial configuration of partons--two events of different initial configurations can give identical $\varepsilon_{2}$. This may cause fluctuations in $v_{2}$, if $v_{2}$ is sensitive to the initial configuration, not simply $\varepsilon_{2}$. We define this part of fluctuations as $\sigma_{v_{2}}\{\rm cfg\}$.  The pentagrams in Fig. 1(c,f) show the averaged $v_{2}$ with identical $\varepsilon_{2}$ from two different sets of events. Each set consists of 3000 events, starting from an identical configuration of initial partons. That is, AMPT starts with exactly identical parton configuration (events have the same initialization), and then evolves with different random number seeds.  The different average $\langle v_{2} \rangle$'s demonstrate that the fluctuations observed in Fig. 1 are not due to statistical fluctuations only; Fluctuations in the initial conditions are important.

\begin{figure*}[htb]
\vskip -0.4cm
\centerline{\includegraphics[width=0.85\textwidth]{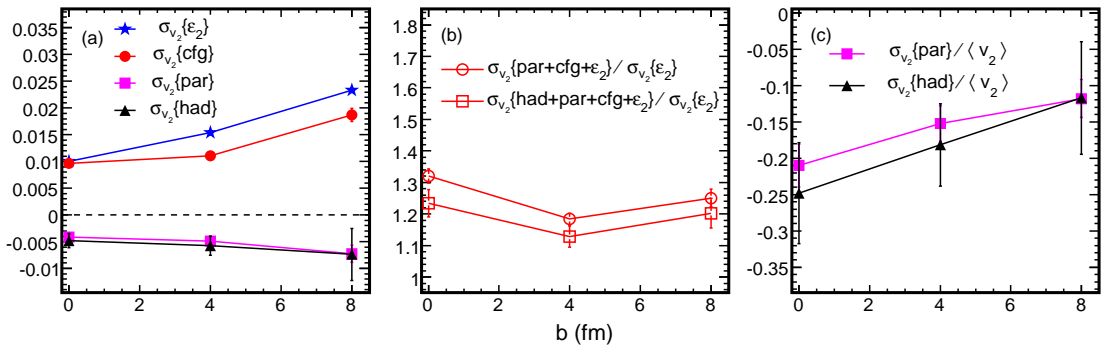}}
\vspace{-1em}
\caption{(Color online) Impact parameter dependence of (a) various $v_{2}$ fluctuations, (b) ratio of total $v_{2}$ fluctuations to that due to eccentricity fluctuations, and (c) various $v_{2}$ fluctuations relative to averaged $\langle v_{2} \rangle$ in Au+Au collisions at $\sqrt{s_{NN}}$ = 200 GeV by the AMPT model.}
\vspace{-1em}
\end{figure*}

The spread of $v_{2}$ in Fig. 1 at a given $\varepsilon_{2}$ is made up from spreads of event-by-event $v_{2}$'s of the different initial configurations about the corresponding average $\langle v_{2} \rangle$'s. The spread of $v_{2}$ for a fixed initial condition comes from statistical fluctuations and possibly dynamical fluctuations. One source of such dynamical fluctuations can be quantum fluctuations in parton interactions. We define this part of fluctuations as $\sigma_{v_{2}}\{\rm par\}$. In case of hadron $v_{2}$, additional fluctuations can arise from hadronization and final-state hadronic scattering processes. We define this part of fluctuations as $\sigma_{v_{2}}\{\rm had\}$.

We now study quantitatively $v_{2}$ fluctuations from the various sources. AMPT is run with default settings and with fixed initial condition. The $v_{2}$ fluctuations of partons and hadrons are obtained from these two ways of running. The fluctuations from the default setting run are $\sigma_{v_{2}}\{\rm par+\rm cfg+\rm sta_{\rm parton}\}$ and $\sigma_{v_{2}}\{\rm had+\rm par+\rm cfg+\rm sta_{\rm hadron}\}$ for partons and hadrons, respectively, and those from the fixed initial-condition run are $\sigma_{v_{2}}\{\rm par+\rm sta_{\rm parton}\}$ and $\sigma_{v_{2}}\{\rm had+\rm par+\rm sta_{\rm hadron}\}$.  The fluctuations are calculated by
\begin{equation}
\sigma_{v_{2}} = \sqrt{\langle v_{2}^{2} \rangle - \langle v_{2} \rangle^{2}}
\end{equation}
where $v_{2}$ is the magnitude of elliptic flow in a single event, given by Eq. (2) and $\langle ... \rangle$ indicates an average over all events at a chosen $\varepsilon_{2}$.

Figure 2 upper panels show the $v_{2}$ fluctuations as a function of $\varepsilon_{2}$. In order to obtain dynamical fluctuations, statistical fluctuations due to finite multiplicities need to be subtracted. The statistical fluctuations of $v_{2}$ are given by
\begin{equation}
\sigma_{v_{2}}\{\rm sta\} = \sqrt{\frac{\langle \cos^{2}(2\phi)\rangle - \langle \cos(2\phi) \rangle^{2}}{N}} = \sqrt{\frac{1 - 2\langle v_{2} \rangle^{2}}{2N}}
\end{equation} This is verified by a Monte Carlo toy model where $N$ particles are generated with $\phi$ angles between 0 and 2$\pi$ according to a $v_{2}$ modulation. We also use the AMPT data themselves to obtain the statistical fluctuation effect by randomly discarding various fractions of particles. The $v_{2}$ fluctuations of the remaining fraction ($f$) of particles are fit to the functional form of $\sqrt{W_{\rm dyn}+\sigma_{v_{2}}^2\{\rm sta\}/\it f}$, where $W_{\rm dyn}$ and $\sigma_{v_{2}}\{\rm sta\}$ are two free parameters. The fitted $\sigma_{v_{2}}\{\rm sta\}$ is found to be consistent with Eq. (5). We subtract the statistical fluctuations given by Eq. (5) from the data in Fig. 2 upper panels. The resulting dynamical fluctuations are shown in the lower panels of Fig. 2. It is found that the dynamical fluctuations for the fixed initial condition data, the $W_{\rm dyn}\{\rm par\}$ and $W_{\rm dyn}\{\rm had+\rm par\}$ in the fit function mentioned above, are both negative. And the latter is larger than the former in terms of magnitude. This suggests that parton interactions and hadronization+hadronic scatterings both introduce negative dynamical fluctuations. The reason may be because these processes tend to fuse particles into fewer ones whereas multi-particle production from single parent tends to introduce additional, i.e. positive, fluctuations. In the following we denote these negative fluctuations as $\sigma_{v_{2}}\{\rm par\}$ = $-\sqrt{-W_{\rm dyn}\{\rm par\}}$ and $\sigma_{v_{2}}\{\rm par+\rm had\}$ = $-\sqrt{-W_{\rm dyn}\{\rm had+\rm par\}}$.

As seen in Fig. 2, the fluctuations, while impact parameter dependent, are approximately independent of $\varepsilon_{2}$ at a given $b$. This is true for both statistical fluctuations subtracted and un-subtracted results (the multiplicities are found to be insensitive to the event-by-event $\varepsilon_{2}$ at a given $b$). We fit the results in the lower panels of Fig. 2 to a constant at each $b$. From the fitted values, we obtain the individual components of $v_{2}$ fluctuations by assuming the different sources of fluctuations are independent of each other, namely
\begin{equation}
\begin{split}
\sigma_{v_{2}}^{2}\{a+b\} = \sigma_{v_{2}}^{2}\{a\}+\sigma_{v_{2}}^{2}\{b\}
 \end{split}
\end{equation}
where $a$ and $b$ stand for two independent fluctuation sources.  There are redundancies in the data in Fig. 2. For example, one can obtain $\sigma_{v_{2}}\{\rm cfg\}$ by taking the difference either between $\sigma_{v_{2}}\{\rm par+\rm cfg\}$ and $\sigma_{v_{2}}\{\rm par\}$ or between $\sigma_{v_{2}}\{\rm had+\rm par+\rm cfg\}$ and $\sigma_{v_{2}}\{\rm had+\rm par\}$. They give consistent results.

Figure 3(a) shows different components of $v_{2}$ fluctuations as a function of $b$. The contributions from eccentricity fluctuations, $\sigma_{v_{2}}\{\varepsilon_{2}\}$ = $k \times \sigma_{\varepsilon_{2}}$, are also shown, where the conversion power (multiplicative factor $k$) from $\varepsilon_{2}$ to $v_{2}$ is obtained from the fits in Fig. 1.  The $\sigma_{v_{2}}\{\rm cfg\}$ and $\sigma_{v_{2}}\{\varepsilon_{2}\}$ are found to increase with increasing $b$. The increase is nearly equally strong. The $\sigma_{v_{2}}\{\rm par\}$ and $\sigma_{v_{2}}\{\rm had\}$, both negative, have weak dependence on $b$. Figure 3(b) shows the ratio of total $v_{2}$ fluctuations to $\sigma_{v_{2}}\{\varepsilon_{2}\}$, the contribution from eccentricity fluctuations. Clearly, the total $v_{2}$ fluctuations in AMPT are larger than eccentricity fluctuations, and there does not seem to have a simple scaling between eccentricity fluctuations and total $v_{2}$ fluctuations. Figure 3(c) shows the negative dynamic fluctuations relative to $\langle v_{2} \rangle$ as a function of $b$. The magnitudes decrease (smaller absolute value) with increasing $b$, which is qualitatively consistent with weaker interactions at larger $b$ that yield smaller negative relative fluctuations.

It is a common perception that $v_{2}$ scales with $\varepsilon_{2}$ and $v_{2}$ fluctuations are dominated by $\varepsilon_{2}$ fluctuations. Our study shows, within the framework of AMPT, that this may not be correct. Those other $v_{2}$ fluctuations can be important and they do not seem to scale with eccentricity fluctuations.

Hydrodynamical calculations of $v_{2}$ do not have any other fluctuations except initial geometry fluctuations (the sum of eccentricity fluctuations and initial configuration fluctuations). If our conclusion is correct that parton interactions, hadronization and final-state hadronic interactions introduce a negative dynamical $v_{2}$ fluctuation effect, and such an effect is relevant in real collision data, then hydrodynamics should have overpredicted $v_{2}$ data. Recently, fluctuations in hydrodynamics, governed by the viscosities, have been investigated~\cite{Kapusta} and are shown to be important and affect the elliptic flow fluctuations besides the initial state fluctuations. This source of fluctuations may be similar in nature to that in parton interactions studied in this work.

{\bf Conclusions}
Elliptic flow and fluctuations are studied by the AMPT model with string melting at three fixed impact parameters in Au+Au collisions at $\sqrt{s_{NN}}$ = 200 GeV.  Both $v_{2}$ of partons and hadrons with respect to the initial participant plane are studied. The average $v_{2}$ is linearly correlated with the average $\varepsilon_{2}$. There is a wide dispersion in $v_{2}$ for a given $\varepsilon_{2}$;  $v_{2}$ is not solely determined by $\varepsilon_{2}$.  Several dynamical fluctuation sources are identified: initial configuration fluctuations at fixed $\varepsilon_{2}$, quantum fluctuations in parton-parton interactions, and those in  hadronization and hadronic scatterings. The fluctuations are studied quantitatively by comparing the $v_{2}$ fluctuations from default settings and with identical parton configuration in AMPT after subtraction of statistical fluctuations. The configuration fluctuations appear independent of $\varepsilon_{2}$ for a given $b$, and increases with increasing $b$. The dynamical fluctuations in $v_{2}$ from parton interactions and hadronization and hadronic scatterings are found to be negative; they reduce $v_{2}$ fluctuations. The total $v_{2}$ fluctuations are larger than the eccentricity fluctuations and they do not seem to scale.

Hydrodynamical models have been very successful in describing experimental data. Hydrodynamics are deterministic; Given an initial configuration space distribution, the final-state momentum anisotropy is fixed. However, our study suggests that there can be additional sources of fluctuations that hydrodynamics do not take into account. Those other fluctuations do not seem to scale with eccentricity fluctuations. This is especially puzzling because experimental data seem to be well described by eccentricity scaling and hydrodynamics. Our finding warrants further investigation of the physics mechanisms of anisotropic flow fluctuations.
	
We have studied AMPT events at fixed impact parameters. It will be interesting to investigate fluctuations as a function of event multiplicity. We have studied $v_{2}$ in this work. It will be interesting to investigate $v_{3}$ fluctuations. We leave such studies to future work.


{\bf Acknowledgments}
This work is supported in part by the National Natural Science Foundation of China under grant No. 11228513, 11221504, 11075060 and 11135011, U.S. Department of Energy under Grant No. DE-FG02-88ER40412, and CCNU-QLPL Innovation Fund (QLPL2011P01) and excellent doctorial dissertation cultivation grant from Central China Normal University.

%

%
\end{document}